\documentclass[11pt]{article}
\pdfoutput=1
\usepackage{palatcm, color, verbatim}
\usepackage{graphicx,graphics, epsfig, psfrag, subfigure, epstopdf}
\usepackage{algorithm, algorithmic, calc,  url, array}
\usepackage{amssymb, amsmath}
\usepackage{multicol,balance}

\begin{document}
\newtheorem{theorem}{Theorem}
\newtheorem{corollary}{Corollary}
\newtheorem{definition}{Definition}
\newtheorem{lemma}{Lemma}

\newcommand{\define}{\stackrel{\triangle}{=}}

\def\QED{\mbox{\rule[0pt]{1.5ex}{1.5ex}}}
\def\proof{\noindent\hspace{2em}{\it Proof: }}

\date{}
\title{Retrospective Interference Alignment} 
\author{\normalsize Hamed Maleki, Syed A. Jafar, Shlomo Shamai\\
       }
%% Notes
\maketitle

\thispagestyle{empty}

%%%%%%%%%%%%%%%%%%%% Abstract %%%%%%%%%%%%%%%%%%%%%%%%%
\begin{abstract}
We explore similarities and differences in recent works on blind interference alignment under different models such as staggered block fading model and the delayed CSIT model. In particular we explore the possibility of achieving interference alignment with delayed CSIT when the transmitters are distributed. Our  main contribution is an interference alignment scheme, called retrospective interference alignment in this work, that is specialized to settings with distributed transmitters. With this scheme we show that the 2 user X channel with only delayed channel state information at the transmitters can achieve 8/7 DoF, while the interference channel with 3 users is able to achieve 9/8 DoF. We also consider another setting where delayed channel output feedback is available to transmitters. In this setting the X channel and the 3 user interference channel are shown to achieve 4/3 and 6/5 DoF, respectively.
\end{abstract}

%%%%%%%%%%%%%%%%%%%% Introduction %%%%%%%%%%%%%%%%%%%%%%%%%
\section{Introduction}
There is much interest in studying the degrees of freedom (DoF) -- and thereby exploring the potential for interference alignment --  in wireless networks in the absence of instantaneous channel state information at the transmitters (CSIT) \cite{Caire_Shamai_ITrans, Jafar_scalar, Jafar_mobile, Lapidoth_Shamai_Wigger_BC, Weingarten_Shamai_Kramer, Gou_Jafar_Wang, Maddah_Compound, Huang_Jafar_Shamai_Vishwanath, Varanasi_noCSIT, Guo_noCSIT, Guo_isotropic, Jafar_corr, Wang_Gou_Jafar, Maddah_Tse}. On the one hand, there are pessmistic results that include \cite{Caire_Shamai_ITrans, Jafar_scalar, Jafar_mobile,  Lapidoth_Shamai_Wigger_BC, Huang_Jafar_Shamai_Vishwanath, Varanasi_noCSIT, Guo_noCSIT, Guo_isotropic} where the DoF are found to collapse due to the inability of the transmitters to resolve spatial dimensions. On the other hand, there are more recent optimistic results \cite{Weingarten_Shamai_Kramer, Gou_Jafar_Wang, Maddah_Compound, Jafar_corr, Wang_Gou_Jafar, Maddah_Tse} where the feasibility of interference alignment is demonstrated under various models of channel uncertainty at the transmitter(s). Closely related to this work are the papers on blind interference alignment \cite{Jafar_corr, Wang_Gou_Jafar} and especially the recent work on interference alignment with delayed CSIT \cite{Maddah_Tse}. 

Reference \cite{Jafar_corr} assumes block fading channels where the coherence blocks are staggered due to difference in coherence times between users. The channels stay constant within a coherence block and are assumed to change instantly across coherence block boundaries. The transmitter(s) have no knowledge of instantaneous channel coefficient values but are assumed to be aware of the coherence block structure of all users. The surprising finding of \cite{Jafar_corr} is the feasibility of alignment based on just the knowledge of the differences in the channel coherence structure across users. \cite{Wang_Gou_Jafar} applies the same principles under a different model where some of the nodes are equipped with reconfigurable antennas capable of switching their own channel states at predetermined time instants, thus allowing further flexibility in determining the channel coherence structure. With this added flexibility \cite{Wang_Gou_Jafar} show that X networks, even with no knowledge of channel coefficient values, do not lose DoF relative to the perfect CSIT setting. The key insight from these blind alignment schemes was that the commonly used i.i.d. fading model is not sufficient for studying the capacity limits of wireless networks even in the absence of CSIT -- because the knowledge of even relatively long term channel statistics can be exploited to achieve interference alignment.

More recently, reference \cite{Maddah_Tse} has introduced the delayed CSIT model, that will also be the main focus of this paper. The delayed CSIT model assumes i.i.d. fading channel conditions, with no knowledge of current channel state at the transmitter. However, perfect knowledge of channel states is available to the transmitter with some delay.  The surprising finding of \cite{Maddah_Tse}, in the context of the vector broadcast (BC) channel, is that not only is CSIT helpful even when it is outdated, but also that it can have a very significant impact as it is capable of increasing the DoF. The delayed CSIT model studied in \cite{Maddah_Tse} is particularly relevant in practice where invariably a delay is involved in any feedback from the receivers to the transmitters. Several recent works point out that channel state information (CSI) can be estimated in principle with sufficient accuracy (estimation error scaling as O$(SNR^{-1})$) to enable the DoF results as SNR becomes large \cite{Jindal, Caire_Jindal_Shamai}. The main obstacle, from a practical perspective, has been the perceived necessity of delivering this CSI to the transmitter before it becomes outdated. The delayed CSIT model therefore opens a  practically meaningful direction to explore the benefits of interference alignment. However, it is one of many possible forms that (delayed) feedback can take in a wireless network. 

The terminology for three closely related delayed feedback models is delineated below through a simple example, for ease of reference in the sequel.
\subsubsection*{Delayed Feedback Models}
Consider a simple Gaussian channel:
\begin{eqnarray}
Y=HX+N
\end{eqnarray}
where $X, Y, N, H$ are the transmitted symbol, the channel output symbol, the additive noise and the i.i.d. time-varying channel, respectively. Perfect channel state information at the receiver (CSIR) is modeled by the assumption that in addition to the channel output symbol $Y$, the receiver also receives the instantaneous channel state $H$ over each channel use. By \emph{delayed} feedback, what is meant is that the information being made available to the transmitter through the feedback channel is based only on past observations at the receivers and, in particular, is independent of the current channel state. Three different settings may be considered.
\begin{enumerate}
\item {\bf Delayed CSIT:} This is the setting where the feedback provides the transmitters only the values of the past channel states $H$ but not the output signals.
\item {\bf Delayed Output Feedback:} This is the setting where the feedback provides the transmitters only the past received signals $Y$ seen by the receivers, but not the channel states explicitly.
\item{\bf Delayed Shannon Feedback:} This is the setting where the feedback provides the transmitters both the past received signals $Y$ as well as the past channel states $H$.
\end{enumerate}

The definitions extend naturally to multiuser settings, although the amount of feedback, and the possible associations between transmitters and receivers on the feedback channel may give rise to many special cases. Clearly, delayed Shannon feedback is the strongest delayed feedback setting, i.e., it can be weakened to obtain either delayed CSIT or the delayed output feedback model by discarding some of the fed back information. Between delayed CSIT and delayed output feedback, neither is a weakened form of the other because in general the output signals $Y$ (even while discounting noise) cannot be inferred from the knowledge of channel states $H$ (e.g., when there is more than one transmitter), and the channel states $H$ cannot be deduced in general from the channel outputs $Y$ even if these are noiselessly made available to the transmitter (e.g., when there are more transmit antennas than receive antennas). In this work, we will be concerned primarily with the delayed CSIT model, and to a lesser extent, with delayed output feedback. But first, we start with a broader review of the similarities and differences between the interference alignment schemes used for blind interference alignment and those used for the Delayed Feedback model.

\section{Similarities between Blind Interference Alignment and Interference Alignment with Delayed CSIT}
While the channel models studied in \cite{Jafar_corr,Wang_Gou_Jafar} and \cite{Maddah_Tse} are quite different, there are some remarkable \emph{essential} similarities in the achievable schemes used in both works that are further expanded in the present work. We start by pointing out the similar aspects of \cite{Jafar_corr, Wang_Gou_Jafar, Maddah_Tse} through a few examples before proceeding to the main contribution of this work.

\subsection{Vector BC with no instantaneous CSIT}
\begin{figure}[!h]
\centering
\includegraphics[width=2.5in]{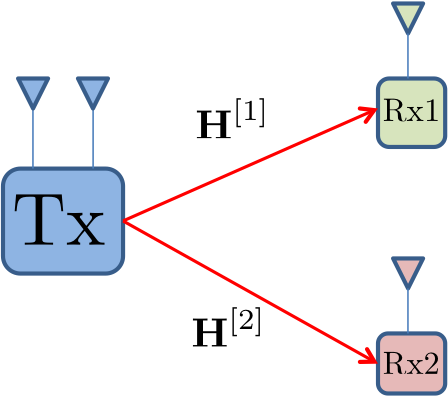}
\caption{Vector BC with No Instantaneous CSIT}
\label{fig:X}
\end{figure}
Consider the broadcast channel (BC) with two single-antenna users, where the transmitter is equipped with two antennas. It is well known that with perfect channel knowledge this channel has $2$ DoF, which can be achieved quite simply by zero forcing. However, the DoF are unknown with partial/limited CSIT for most cases and it is believed that no more than 1 DoF may be achievable in general. An outer bound of $\frac{4}{3}$ DoF has been shown to be applicable to a wide variety of limited CSIT scenarios \cite{Lapidoth_Shamai_Wigger_BC,Weingarten_Shamai_Kramer, Maddah_Tse}. In particular, \cite{Maddah_Tse} explicitly shows the $\frac{4}{3}$ outer bound for delayed CSIT. It is easily verified that the outer bound applies under the stronger setting of delayed Shannon feedback as well. Further the same outer bound applies to the staggered block fading model. Recently, reference \cite{Jafar_corr} shows that $\frac{4}{3}$ DoF are achievable under a staggered block fading model where the two users have staggered coherent fading blocks. Reference \cite{Maddah_Tse} shows the achievability of $\frac{4}{3}$ DoF under the assumption that only delayed CSIT is available. In both cases the achievable scheme is described as follows:
\begin{itemize}
\item Use the channel three times to send two information symbols to each user.
\item  In one time slot, two symbols for user 1 are sent simultaneously from the two transmit antennas, providing user 1 one linear combination of the two desired symbols. One more linear combination would be needed to resolve the desired symbols at user 1. User 2 simultaneously also obtains a linear combination of these undesired symbols, but does not need to resolve them.
\item In another time slot, two symbols for user 2 are sent simultaneously from the two transmit antennas, thus similarly providing user 2 one linear combination of his desired symbols, and providing user 1 a linear combination of these undesired symbols.
\item In the final transmission, both users are simultaneously provided another linear combination of their respective desired symbols. The key to alleviate interference in this time slot is that the linear combination of undesired symbols received by a user in the third time slot is \emph{the same linear combination} that he already received earlier, which allows the receiver to cancel the interference and then recover his desired symbols. 
\end{itemize}
The key to both schemes is the third time slot. Because of different channel models, the manner in which the third time slot transmission is accomplished is different in \cite{Jafar_corr} and \cite{Maddah_Tse}. In \cite{Jafar_corr} the staggered coherence times ensure that each user receives the final transmission over a channel state that is identical to his channel state when he  received interference in a prior time slot, but different from his channel state where he previously received his desired symbols. The channel state determines the linear combination and thus, the desired symbols are seen twice with different linear combinations while the interfering symbols are seen twice in the same linear combination, allowing interference cancellation. In \cite{Maddah_Tse} the same effect is accomplished by using delayed CSIT feedback. In the third time slot -- because the transmitter now knows the channel states from the previous two time slots -- the transmitter is able to send (from only one antenna) a superposition of the  linear combinations of the undesired symbols seen previously by the two users. Since these linear combinations were received before, undesired information is easily cancelled, leaving only the linear combination of desired symbols that provides the second equation in order to solve for the two desired variables. Since each user is able to resolve his 2 desired symbols over a total of 3 time slots, the DoF of $\frac{4}{3}$ are achieved.

\subsection{Vector BC with instantaneous CSIT for User 1}
\begin{figure}[!h]
\centering
\includegraphics[width=2.5in]{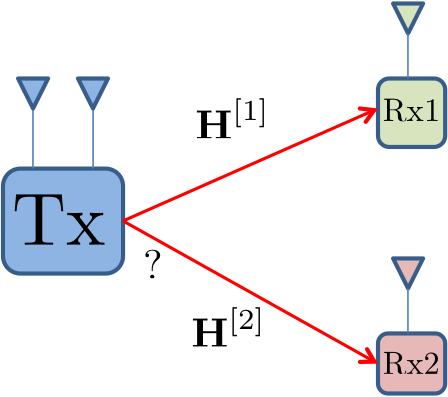}
\caption{Vector BC with Instantaneous CSIT for User 1}
\label{fig:X}
\end{figure}Consider the same vector BC as before, with the difference that now we assume the channel state of User 1 is instantaneously available to the transmitter. The DoF outer bound in this case is $\frac{3}{2}$ \cite{Weingarten_Shamai_Kramer} and is also applicable to a broad class of channel uncertainty models for User 2. In particular, it can be shown to be applicable to the delayed Shannon feedback model, by providing all the information available to Receiver 2 also to Receiver 1, thus making it a physically degraded broadcast channel for which it is known that feedback does not increase capacity \cite{ElGamal_FB}. Then, proceeding without feedback, the outer bound arguments for the compound setting in \cite{Weingarten_Shamai_Kramer} can be extended to this setting, producing the same DoF outer bound of $\frac{3}{2}$. Note that an outer bound for delayed Shannon feedback is also an outer bound for delayed CSIT. Similar arguments (without feedback) are applicable to show the $\frac{3}{2}$ DoF outer bound for blind interference alignment (staggered block fading) model as well.  Further, the achievability of $\frac{3}{2}$ DoF can be shown under both settings of staggered block fading and delayed CSIT in essentially the same manner, as described below.
\begin{itemize}
\item Use the channel twice to send two information symbols to User 1 and one information symbol to User 2.
\item User 2's information symbol is sent along a beamforming vector orthogonal to User 1's known channel vector, so User 1 sees no interference due to User 2.
\item User 1's two symbols are sent twice in a manner that the same  linear combination of the two undesired symbols is experienced by User  2 in both timeslots. This allows User 2 to cancel interference from one of the two time slots to recover his one desired symbol. However, User 1 sees two different linear combinations of his desired symbols and no interference, so he is able to recover both symbols.
\end{itemize}
Thus, 1 DoF is achieved by User 1 and 0.5 DoF by User 2, for a total of 1.5 DoF which is also the outer bound.
The key here is to transmit the same linear combination of undesired symbols for User 2 while User 1 sees two different linear combinations. \cite{Jafar_corr} does this by assuming that the channel stays constant for User 2 while it changes for User 1. The delayed CSIT setting can also accomplish the same effect because once the transmitter learns the channel states, it knows the linear combination of undesired symbols already seen by User 2 and re-sends the same linear combination from one antenna. Since this linear combination has not yet been seen by User 1, it gives him the second equation he needs, while for User 2 it is  just a repetition of the previously seen interference which can be cancelled. Interestingly, in both cases (staggered block fading or delayed CSIT) the channel knowledge of User 1 (i.e., the user with the known channel) is only needed by the transmitter for one of the two channel uses (because User 2's symbol can be transmitted only once \cite{Jafar_corr}).

\subsubsection{The (1,2,3,4) Two User MIMO Interference Channel}
\begin{figure}[!h]
\centering
\includegraphics[width=3in]{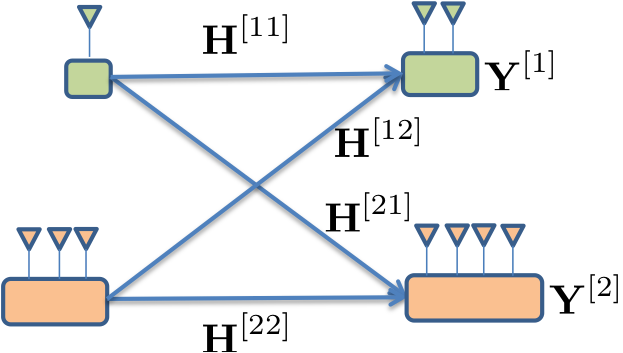}
\caption{(1,2,3,4) MIMO IC}
\label{fig:MIMOIC}
\end{figure}
Shown in Fig. \ref{fig:MIMOIC}, the (1,2,3,4) MIMO IC is a two user interference channel where User 1 has 1 transmit and 2 receive antennas, while User 2 has 3 transmit and 4 receive antennas. This particular channel configuration was first highlighted in \cite{Huang_Jafar_Shamai_Vishwanath} as an example where interference alignment was a possibility. Specifically, the question is, what is the maximum DoF possible for User 2 if User 1 simultaneously achieves his maximum value of 1 DoF? With no interference alignment the result would be only 1 DoF for user 2, but with interference alignment it may be possible to achieve up to 1.5 DoF for User 2. While recent work in \cite{Guo_noCSIT} showed that under i.i.d. isotropic fading, User 2 cannot achieve more than 1 DoF, thus precluding the possibility of interference alignment, the question remains open for non-iid, non-isotropic models. In particular, \cite{Jafar_corr} shows that under a staggered coherence block fading model, User 2 may indeed achieve 1.5 DoF at the same time that User 1 achieves his maximum DoF =1. Here we briefly summarize how this result is shown in \cite{Jafar_corr} and how the same scheme can be translated to the delayed CSIT setting (although it remains to be verified whether the same outer bound applies to delayed CSIT setting for the (1,2,3,4) MIMO IC), once again highlighting the similarity of the two.
\begin{itemize}
\item Operate the channel over two time slots, sending a new symbol in each time slot from Transmitter 1 and sending three symbols from Transmitter 2, repeated over two slots.
\item The key is to send the three symbols from Transmitter 2 in such a manner that User 1, at his two receive antennas, sees the same two linear combinations of undesired symbols twice. Thus, in each time slot he is able to free up one dimension by cancelling the corresponding interference using the same linear combination received in the other time slot, and recover his desired symbol.
\end{itemize}
In order to make sure that the same two linear combinations of undesired symbols are seen on the two receive dimensions twice by Receiver 1, reference\cite{Jafar_corr} exploits the assumption that the channel between Transmitter 2 and Receiver 1 stays constant over two time slots. However, in the delayed CSIT setting, after the first time slot Transmitter 2 knows the CSIT from the first time slot and therefore also knows the two different linear combinations of the 3 transmitted symbols observed at the two receive antennas of Receiver 1. In time slot 2 then Transmitter 2 sends the two linear combinations, each from a different transmit antenna.  Since only $2$ antennas are used in the second time slot, the $2\times 2$channel can be inverted by Receiver 1 to essentially experience the same two linear combinations of undesired symbols on the two resulting receive antennas. Thus, one receive antenna can be cleared of interference in each time slot by using the corresponding linear combination of undesired symbols observed in the other time slot, allowing the desired symbol to be resolved in each time slot.

\section{Differences Between Staggered Fading and Delayed CSIT Settings}

In spite of the strong similarities between the staggered fading model \cite{Jafar_corr} and the delayed CSIT \cite{Maddah_Tse} model, the two settings have very marked differences in general. The difference becomes evident as soon as the question of alignment of signals from \emph{different} transmitters comes to light. Note that in all examples described above the signals being aligned are from the same transmitter. However, when we go to more distributed settings, e.g., the X or $K$ user interference channels the difference between the two settings is quite stark. With \emph{suitably} staggered coherent times/antenna switching, it is shown in \cite{Jafar_corr, Wang_Gou_Jafar} that neither the $X$ network with $M$ transmitters and $N$ receivers, nor the $K$ user interference channel lose \emph{any} DoF relative to the setting where perfect, global and instantaneous CSIT is available. The $M\times N$ user $X$ network can still achieve $\frac{MN}{M+N-1}$ DoF, and the $K$ user interference channel can still achieve $\frac{K}{2}$ DoF even with no instantaneous (or delayed) CSIT besides the knowledge of the channel coherence structure/antenna switching pattern. However, in the delayed CSIT setting, (albeit only with i.i.d fading) the outer bounds clearly show a loss in DoF. Thus, evidently there is no complete theoretical equivalence between the two settings. 

The significance of distributed transmitters brings us to the issue central to the present work. We know that in the delayed CSIT setting, DoF are lost because the DoF outer bound for the vector BC (say with $K$ transmit antennas, and $K$ single antenna users) loses to the $\frac{K}{2}$ DoF that are achievable in the $K$ user interference channel both in the compound setting (with finite number of generic states \cite{Gou_Jafar_Wang}) as well as with suitable staggered coherence block model. However, a very interesting question remains open -- Does transmitter cooperation improve DoF even with channel uncertainty at transmitters?

The main difference between the $X$ channel and the vector BC is that in the vector BC  the transmitters are allowed to cooperate as multiple antennas of one transmit node. With full CSIT it is well known that the vector BC has larger DoF than the X channel, so the cooperation between transmitters increases DoF. However, with limited CSIT, e.g. in the compound channel setting \cite{Weingarten_Shamai_Kramer, Gou_Jafar_Wang, Maddah_Compound}, somewhat surprisingly, several recent results have shown that the DoF are the same with and without transmitter cooperation, i.e., whether it is the X channel or the corresponding vector BC. Indeed, with the staggered fading model of \cite{Jafar_corr}, the same approach works for the X channel as does for the vector BC. 

The delayed CSIT setting gives us another framework within which one can hope to gain additional insights into the role of transmitter cooperation in determining the available DoF under channel uncertainty.  Consider the vector BC with 2 antennas at the transmitter and 2 single-antenna users, for which we know DoF = $\frac{4}{3}$ with delayed CSIT. However, does the same scheme work for the $X$ channel? More generally, what DoF can we achieve on the $X$ channel with delayed CSIT? These are the questions that motivate the current work. 

We start with two interesting observations.
\begin{itemize}
\item {\bf The approach used for vector BC in \cite{Maddah_Tse} is not directly applicable to the X channel.} This is  because in the vector BC, after the transmitter acquires delayed CSIT, it is able to reconstruct the linear combinations of undesired symbols previously seen by the receivers. However, in the X channel, even after the distributed transmitters learn the channel states, they cannot re-send the same linear combination of undesired symbols. For example, consider the first time slot where both transmitters send a symbol each intended for User 1, which provides User 2 a particular linear combination of the two undesired symbols. Future repetitions of this particular linear combination carry no interference cost to User 2, because he is able to cancel the interference. However, in the X channel, one cannot repeat this linear combination, because if each transmitter repeats its own symbol meant for User 1, the linear combination seen by User 2 also depends on the \emph{current} channel state which cannot be compensated for by the transmitters who have, at this point, no knowledge of it. This is why in the broadcast setting the repetitions of a particular linear combination take place from a single transmit antenna as a scalar value. For example, if the transmitter wants User 2 to see the linear combination $2u_1+u_2$, it cannot transmit $u_1$ and $u_2$ from two different antennas, which would leave the resulting linear combination in the hands of nature. Instead, in the vector BC with delayed CSIT, the transmitter sends $2u_1+u_2$ from the same antenna, so that even when nature scales the transmitted scalar value, the receiver will see $h(2u_1+u_2)$ from which $h$ can be scaled off. The BC can do this because both $u_1, u_2$ symbols are available to e.g. antenna 1. In the $X$ channel on the other hand, transmitter 1 may only know $u_1$ and transmitter $2$ only $u_2$,  thus forcing the two symbols to be sent from different antennas, and since the channels change every time slot (or before the CSIT becomes available), it is not possible to repeat the linear combination seen by User 2\footnote{Note that if instead of delayed CSIT, we have delayed output feedback, then this problem does not arise.}.

\item {\bf Aligning interfering symbols from the same transmitter does not produce DoF benefits on the X channel.} The next natural thought is this -- since, as discussed above, it seems we cannot repeat the alignment of symbols from two different transmitters without knowledge of current channel state, can we instead achieve interference alignment between two undesired symbols coming from the \emph{same} transmitter? This is a subtle but important point. As we argue here, in the X channel, alignment of symbols (with linear beamforming schemes) from the same transmitter  is not beneficial for DoF. The reason is that we are considering the X channel where all nodes have equal number of antennas. Suppose we have two symbols that originate at the same transmitter (say Transmitter 1) and are aligned at a particular receiver (say Receiver 1). Since this alignment does not depend on the other transmitter, let us ignore all signals received from the other transmitter and focus only on the aligned signals received from Transmitter 1. Even with symbol extensions, barring degenerate conditions, what it means is that Receiver 1 can invert the channel to Transmitter 1, and thus observe the transmitted symbols (within the noise distortion level, which is inconsequential for DoF arguments). Thus, if two undesired symbols from the same transmitter are seen aligned at a particular receiver, it must be because they are aligned at the transmitter itself. Now, if these symbols are aligned at the transmitter itself, then they cannot be resolved further downstream at any other receiver, and in particular, at the receiver that wants these two symbols and therefore must not see them aligned together. Thus the benefits of alignment are lost from a DoF perspective if the aligned symbols come from the same transmitter on the X channel.
\end{itemize}

The two observations listed above, seem to leave very little hope of achieving interference alignment on the X channel. If we cannot repeat a linear combination from two different transmitters without current channel knowledge, and as we argue above, it does not help to align interference from the same transmitter, then there are few alternatives left and it seems the DoF of the X channel will be limited to 1. However, as we show in this paper, this is not the case. The key to the positive result is that while the vector BC approach does not directly extend to the X channel, an interesting new approach -- that we call \emph{retrospective interference alignment} -- is able to achieve interference alignment on the X channel even with delayed CSIT.

We proceed to the main result next.

\section{The X Channel}
The X channel consists of two transmitters, two receivers, and four independent messages, one from each transmitter to each receiver. We assume that the channels vary in an i.i.d. fashion according to some continuous distribution with support that is bounded away from zero and infinity. The receivers are assumed to have perfect knowledge of all channel states. The transmitters do not know the current channel state, but they do have access to all channel states up to the previous channel use. This model is referred to as the \emph{delayed CSIT} model. Our goal is to explore the DoF that can be achieved by the X channel in this setting. The definitions of achievable data rates, capacity, power constraints, DoF are all used in the standard sense as in e.g. \cite{Maddah_Tse} and will not be repeated here. We also assume that the reader is familiar with DoF analysis when working with linear beamforming schemes, and in particular the requirements for interference alignment. For these and other standard issues such as -- why we  ignore noise in this analysis, what are the conditions for desired signals to be recovered in the presence of interference, a literature survey of earlier works on interference alignment and DoF such as \cite{Zheng_Tse, Jafar_Shamai, MMK, Cadambe_Jafar_int} is recommended.

\begin{theorem}
The 2 user X channel with delayed CSIT, can achieve DoF = $\frac{8}{7}$ almost surely.
\end{theorem}
\begin{figure}[!h]
\centering
\includegraphics[width=3in]{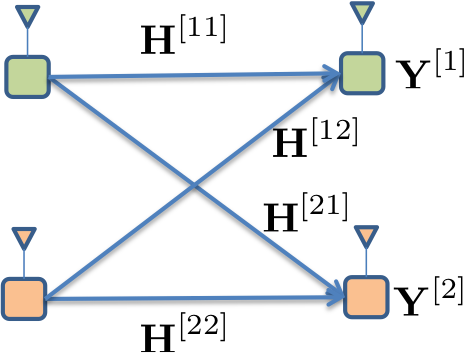}
\caption{X Channel}
\label{fig:X}
\end{figure}
\proof
We wish to show that $\frac{8}{7}$ DoF are achievable on the X channel with delayed CSIT. To this end, consider a 7 symbol extension of the channel, i.e., we will consider 7 channel uses over which the precoding vectors will be designed. Each of the 4 messages $W^{[kj]}$ will be assigned two precoding vectors $\overline{\bf V}^{[kj]}=\left[\overline{\bf V}^{[kj]}_{1} ~~\overline{\bf V}^{[kj]}_{2}\right]$, where $\overline{\bf V}^{[kj]}_{i}$ is the $i^{th}$ column of the $7\times 2$ matrix $\overline{\bf V}^{[kj]}$, $k=1,2$, and will carry the scalar coded information symbol $u^{[kj]}_i$. Thus, over the $7$ symbol block, the received signal at Receiver $k$, $k=1,2$, can be expressed as:
\begin{eqnarray}
\overline{\bf Y}^{[k]}&=&\overline{\bf H}^{[k1]}\left(\overline{\bf V}^{[11]}_{ 1}u^{[11]}_1+\overline{\bf V}^{[11]}_{ 2}u^{[11]}_2+\overline{\bf V}^{[21]}_{ 1}u^{[21]}_1+\overline{\bf V}^{[21]}_{ 2}u^{[21]}_2\right)\nonumber\\
&&~~~~+\overline{\bf H}^{[k2]}\left(\overline{\bf V}^{[12]}_{ 1}u^{[12]}_1+\overline{\bf V}^{[12]}_{ 2}u^{[12]}_2+\overline{\bf V}^{[22]}_{ 1}u^{[22]}_1+\overline{\bf V}^{[22]}_{ 2}u^{[22]}_2\right)+\overline{\bf Z}^{[k]}\\
&=& \overline{\bf H}^{[k1]}\left[ \overline{\bf V}^{[11]}~~~\overline{\bf V}^{[21]}\right]{\bf U}^{[\star 1]}+\overline{\bf H}^{[k2]}\left[ \overline{\bf V}^{[12]}~~~\overline{\bf V}^{[22]}\right]{\bf U}^{[\star 2]}+\overline{\bf Z}^{[k]}
\end{eqnarray}
Here $\overline{\bf Y}^{[k]}, \overline{\bf Z}^{[k]}$ are $7\times 1$ vectors, $\overline{\bf H}^{[kj]}$ are $7\times 7$ diagonal matrices and $u^{[kj]}_i$ are scalar symbols. ${\bf U}^{[\star i]}=\left[u^{[1i]}_1, u^{[1i]}_2, u^{[2i]}_1, u^{[2i]}_2\right]^T$ is the $4\times 1$ vector of information symbols originating at transmitter $i$.

The key task is to design the precoding vectors $\overline{\bf V}^{[kj]}_{ i}$ to achieve interference alignment. The elements of the precoding vector can depend on the channel state only retrospectively, i.e., at time $n$,  $ n= 1,2, \cdots, 7$, the $n^{th}$ element of a precoding vector can depend on the past channel states $\overline{\bf H}^{[\star\star]}(1), \cdots, \overline{\bf H}^{[\star\star]}(n-1)$ but not the present or future channel states.

\subsubsection*{First 3 Channel Uses}
For the first $3$ channel uses each transmitter simply sends random linear combinations (the coefficients are generated randomly offline and shared between all transmitters and receivers before the beginning of communication) of the 4 symbols originating at that transmitter. A different linear combination is sent each time. This gives each receiver $3$ equations in $8$ variables. Because the combining coefficients are chosen randomly and independently of the channel realizations these equations contain no particular structure and may be regarded as \emph{generic} linear equations. Ignoring noise, the received signals at Receiver 1 at this stage can be expressed as:
\begin{allowdisplaybreaks}
\begin{eqnarray}
\left[\begin{array}{c}\overline{\bf Y}^{[1]}(1)\\\overline{\bf Y}^{[1]}(2)\\\overline{\bf Y}^{[1]}(3)\end{array}\right]&=&
 \left[\begin{array}{cccc}
\overline{\bf H}^{[11]}(1)\overline{\bf V}^{[11]}_1(1) & \overline{\bf H}^{[11]}(1)\overline{\bf V}^{[11]}_2(1) &\overline{\bf H}^{[12]}(1)\overline{\bf V}^{[12]}_1(1)&\overline{\bf H}^{[12]}(1)\overline{\bf V}^{[12]}_2(1)\\
\overline{\bf H}^{[11]}(2)\overline{\bf V}^{[11]}_1(2) & \overline{\bf H}^{[11]}(2)\overline{\bf V}^{[11]}_2(2) &\overline{\bf H}^{[12]}(2)\overline{\bf V}^{[12]}_1(2)&\overline{\bf H}^{[12]}(2)\overline{\bf V}^{[12]}_2(2)\\
\overline{\bf H}^{[11]}(3)\overline{\bf V}^{[11]}_1(3) & \overline{\bf H}^{[11]}(3)\overline{\bf V}^{[11]}_2(3) &\overline{\bf H}^{[12]}(3)\overline{\bf V}^{[12]}_1(3)&\overline{\bf H}^{[12]}(3)\overline{\bf V}^{[12]}_2(3)
\end{array}\right]
\left[\begin{array}{c}
u^{[11]}_1\\ u^{[11]}_2\\ u^{[12]}_1\\ u^{[12]}_2
\end{array}\right]\nonumber\\
 &&+\left[\begin{array}{cccc}
\overline{\bf H}^{[11]}(1)\overline{\bf V}^{[21]}_1(1) & \overline{\bf H}^{[11]}(1)\overline{\bf V}^{[21]}_2(1) &\overline{\bf H}^{[12]}(1)\overline{\bf V}^{[22]}_1(1)&\overline{\bf H}^{[12]}(1)\overline{\bf V}^{[22]}_2(1)\\
\overline{\bf H}^{[11]}(2)\overline{\bf V}^{[21]}_1(2) & \overline{\bf H}^{[11]}(2)\overline{\bf V}^{[21]}_2(2) &\overline{\bf H}^{[12]}(2)\overline{\bf V}^{[22]}_1(2)&\overline{\bf H}^{[12]}(2)\overline{\bf V}^{[22]}_2(2)\\
\overline{\bf H}^{[11]}(3)\overline{\bf V}^{[21]}_1(3) & \overline{\bf H}^{[11]}(3)\overline{\bf V}^{[21]}_2(3) &\overline{\bf H}^{[12]}(3)\overline{\bf V}^{[22]}_1(3)&\overline{\bf H}^{[12]}(3)\overline{\bf V}^{[22]}_2(3)
\end{array}\right]
\left[\begin{array}{c}
u^{[21]}_1\\ u^{[21]}_2\\ u^{[22]}_1\\ u^{[22]}_2
\end{array}\right]\nonumber\\
\Rightarrow {\bf Y}^{[1]}&=& \left[{\bf H}^{[11]}{\bf V}^{[11]}_1 ~~~ {\bf H}^{[11]}{\bf V}^{[11]}_2 ~~~ {\bf H}^{[12]}{\bf V}^{[12]}_1 ~~~{\bf H}^{[12]}{\bf V}^{[12]}_2\right]\left[u^{[11]}_1, u^{[11]}_2, u^{[12]}_1, u^{[12]}_2\right]^T\nonumber\\
&&+\left[{\bf H}^{[11]}{\bf V}^{[21]}_1 ~~~ {\bf H}^{[11]}{\bf V}^{[21]}_2 ~~~ {\bf H}^{[12]}{\bf V}^{[22]}_1 ~~~{\bf H}^{[12]}{\bf V}^{[22]}_2\right]\left[u^{[21]}_1, u^{[21]}_2, u^{[22]}_1, u^{[22]}_2\right]^T
\end{eqnarray}
\end{allowdisplaybreaks}
Note that the channel and precoding vectors without the overbar notation refer to the values over only the first 3 channel uses. The received signal for Receiver 2, is also defined similarly.
\begin{eqnarray}
\Rightarrow {\bf Y}^{[2]}&=& \left[{\bf H}^{[21]}{\bf V}^{[11]}_1 ~~~ {\bf H}^{[21]}{\bf V}^{[11]}_2 ~~~ {\bf H}^{[22]}{\bf V}^{[12]}_1 ~~~{\bf H}^{[22]}{\bf V}^{[12]}_2\right]\left[u^{[11]}_1, u^{[11]}_2, u^{[12]}_1, u^{[12]}_2\right]^T\nonumber\\
&&+\left[{\bf H}^{[21]}{\bf V}^{[21]}_1 ~~~ {\bf H}^{[21]}{\bf V}^{[21]}_2 ~~~ {\bf H}^{[22]}{\bf V}^{[22]}_1 ~~~{\bf H}^{[22]}{\bf V}^{[22]}_2\right]\left[u^{[21]}_1, u^{[21]}_2, u^{[22]}_1, u^{[22]}_2\right]^T
\end{eqnarray}
\subsubsection*{Last 4 Channel Uses}
During the last 4 channel uses we will operate over a new effective set of variables. Using the terminology of \cite{Maddah_Tse} we can call these the second layer variables. The goal will be to allow each receiver to resolve all layer-2 variables. Since we have only $4$ channel uses left, and each channel use will provide only one equation to each receiver, we will choose $4$ layer-2 variables.

While everything so far is consistent with the approaches of \cite{Jafar_corr, Maddah_Tse}, at this point our approach goes into a new direction. As mentioned earlier, because the transmitters are distributed, they do not have access to any of the equations available so far to the receivers -- because each equation contains symbols from both transmitters. Thus, unlike the broadcast setting studied in \cite{Maddah_Tse} where the transmit antennas are co-located, we cannot construct layer-2 variables out of the equations already available to the receivers. Instead, we will construct layer-2 variables out of the symbols available at each transmitter. Let us define our new variables:

\begin{eqnarray}
s^{[1]}_1&=&u^{[11]}_1 - \gamma^{[1]}_1 u^{[11]}_2\\
s^{[2]}_1&=&u^{[12]}_1- \gamma^{[2]}_1 u^{[12]}_2\\
s^{[1]}_2&=&u^{[21]}_1- \gamma^{[1]}_2 u^{[21]}_2\\
s^{[2]}_2&=&u^{[22]}_1- \gamma^{[2]}_2 u^{[22]}_2
\end{eqnarray}
where $\gamma^{[k]}_i$ are constants whose values will be specified soon after we arrive at the rationale for choosing these values. Consistent with our causality and delayed CSIT constraint, the constants will depend on the channel values only from phase-I, i.e., from the first 3 channel uses. The most important observation here is that the variables $s^{[k]}_1, s^{[k]}_2$ are available to Transmitter $k$.

Once the layer-2 variables are defined, the operation over the last 4 channel uses is very simple. Each transmitter sends a different linear combination of its two layer-2 variables over each channel use. Each receiver sees 4 different linear combinations of the 4 layer-2 variables (2 from each transmitter) over the 4 channel uses, and is therefore able to resolve each variable (again, ignoring noise -- for DoF arguments).

{\it Remark:} Note that no CSIT, not even delayed CSIT,  is needed for the last 4 channel uses. This is an important observation that could significantly reduce the overhead of feeding back the delayed CSIT to the transmitters. Evidently, no more than $\frac{3}{7}$ of the channel states need to be fed back with the proposed scheme.

\subsubsection*{Retrospective Interference Alignment}
As mentioned above, the novelty of \emph{retrospective interference alignment} lies in the construction of layer-2 variables. In particular, we will choose the values of the constants $\gamma^{[k]}_i$ to align interference over the \emph{first 3 channel uses} -- i.e., acting retrospectively. Also note that the definitions of layer-2 variables seem to suggest at first that the variables from the same transmitter are being aligned into the layer-2 variables. This is not the correct interpretation, for the simple reason that alignment of variables from the same transmitter through linear schemes cannot provide DoF benefits on the SISO X channel, as argued earlier. As we show below, the actual alignment still happens between information variables coming from different transmitters over the first 3 channel uses. 

From phase-2 we know that both users are able to solve for the layer-2 variables. Now, let us consider the $3$ equations accumulated at each receiver over the first 3 channel uses in 8 variables. Let us substitute for four of these variables in terms of the solved values of the layer-2 variables. Specifically, we make the following substitutions:
\begin{eqnarray}
u^{[11]}_1 &\longrightarrow& s^{[1]}_1+\gamma^{[1]}_1 u^{[11]}_2\\
u^{[12]}_1 &\longrightarrow& s^{[2]}_1+\gamma^{[2]}_1 u^{[12]}_2\\
u^{[21]}_1 &\longrightarrow& s^{[1]}_2+\gamma^{[1]}_2 u^{[21]}_2\\
u^{[22]}_1 &\longrightarrow& s^{[2]}_2+\gamma^{[2]}_2 u^{[22]}_2
\end{eqnarray}
Note that after these substitutions there are only four unknown variables left (since the $s^{[k]}_i$ are already known from Phase-2) --- $u^{[11]}_2, u^{[12]}_2, u^{[21]}_2, u^{[22]}_2$. Since we have four variables and only three equations we will need interference alignment. Out of the 4 remaining unknown variables each receiver only desires 2. As usual on the  X channel, the 2 undesired variables will be aligned into one dimension, leaving the remaining two dimensions available to recover the two desired variables. This alignment will be enabled precisely by the choice of the $\gamma^{[k]}_i$ in the layer-2 variable definitions -- thus accomplishing retrospective interference alignment.

Following the substitutions, consider Receiver 1, where we have three equations in the remaining 4 variables (Note that the following quantities -- without overbar notation -- refer to only the first three channel uses).
\begin{eqnarray}
{\bf Y}^{[1]}-{\bf H}^{[11]}{\bf V}^{[11]}_1s^{[1]}_1-{\bf H}^{[12]}{\bf V}^{[12]}_1s^{[2]}_1-{\bf H}^{[11]}{\bf V}^{[21]}_1s^{[1]}_2-{\bf H}^{[12]}{\bf V}^{[22]}_1s^{[2]}_2=\nonumber\\
{\bf H}^{[11]}{\bf V}^{[11]}_1\gamma^{[1]}_1u^{[11]}_2+ {\bf H}^{[11]}{\bf V}^{[11]}_2 u^{[11]}_2+{\bf H}^{[12]}{\bf V}^{[12]}_1\gamma^{[2]}_1u^{[12]}_2+{\bf H}^{[12]}{\bf V}^{[12]}_2u^{[12]}_2\nonumber\\
{\bf H}^{[11]}{\bf V}^{[21]}_1\gamma^{[1]}_2u^{[21]}_2+ {\bf H}^{[11]}{\bf V}^{[21]}_2 u^{[21]}_2+{\bf H}^{[12]}{\bf V}^{[22]}_1\gamma^{[2]}_2u^{[22]}_2+{\bf H}^{[12]}{\bf V}^{[22]}_2u^{[22]}_2
\end{eqnarray}

\noindent The interfering symbols $u^{[21]}_2, u^{[22]}_2$ arrive along directions:
\begin{eqnarray}
u^{[21]}_2&\Rightarrow& {\bf H}^{[11]}{\bf V}^{[21]}_1\gamma^{[1]}_2+ {\bf H}^{[11]}{\bf V}^{[21]}_2 \\
u^{[22]}_2&\Rightarrow&{\bf H}^{[12]}{\bf V}^{[22]}_1\gamma^{[2]}_2+{\bf H}^{[12]}{\bf V}^{[22]}_2
\end{eqnarray}
The RHS of the above expressions are the $3\times 1$ vectors indicating the direction along which interference is seen by Receiver 1 from the two undesired symbols. For interference alignment we would like these directions to be co-linear.
\begin{eqnarray}
{\bf H}^{[11]}{\bf V}^{[21]}_1\gamma^{[1]}_2+ {\bf H}^{[11]}{\bf V}^{[21]}_2 &=&\beta\left({\bf H}^{[12]}{\bf V}^{[22]}_1\gamma^{[2]}_2+{\bf H}^{[12]}{\bf V}^{[22]}_2\right)
\end{eqnarray}
for some constant $\beta$. Equivalently
\begin{eqnarray}
\left[{\bf H}^{[11]}{\bf V}^{[21]}_1 ~~~ {\bf H}^{[11]}{\bf V}^{[21]}_2 ~~~ {\bf H}^{[12]}{\bf V}^{[22]}_1~~~{\bf H}^{[12]}{\bf V}^{[22]}_2\right]_{3\times 4}
\left[\begin{array}{c}
\gamma^{[1]}_2\\
1\\
-\beta\gamma^{[2]}_2\\
-\beta
\end{array}
\right]={\bf 0}_{3\times 1}
\end{eqnarray}
Since the matrix on the left is generic and of size $3\times 4$ it has a unique (upto scaling) null vector. The choice of the values of $\gamma^{[2]}_1, \gamma^{[2]}_2, \beta$ is made precisely to force the vector on the right to be this null vector, thus aligning interference.

Similarly, at Receiver 2, we have three equations in four information symbols:
\begin{eqnarray}
{\bf Y}^{[2]}-{\bf H}^{[21]}{\bf V}^{[11]}_1s^{[1]}_1-{\bf H}^{[22]}{\bf V}^{[12]}_1s^{[2]}_1-{\bf H}^{[21]}{\bf V}^{[21]}_1s^{[1]}_2-{\bf H}^{[22]}{\bf V}^{[22]}_1s^{[2]}_2=\nonumber\\
{\bf H}^{[21]}{\bf V}^{[11]}_1\gamma^{[1]}_1u^{[11]}_2+ {\bf H}^{[21]}{\bf V}^{[11]}_2 u^{[11]}_2+{\bf H}^{[22]}{\bf V}^{[12]}_1\gamma^{[2]}_1u^{[12]}_2+{\bf H}^{[22]}{\bf V}^{[12]}_2u^{[12]}_2\nonumber\\
{\bf H}^{[21]}{\bf V}^{[21]}_1\gamma^{[1]}_2u^{[21]}_2+ {\bf H}^{[21]}{\bf V}^{[21]}_2 u^{[21]}_2+{\bf H}^{[22]}{\bf V}^{[22]}_1\gamma^{[2]}_2u^{[22]}_2+{\bf H}^{[22]}{\bf V}^{[22]}_2u^{[22]}_2
\end{eqnarray}
and the interfering symbols $u^{[11]}_2, u^{[12]}_2$ arrive along the directions:
\begin{eqnarray}
u^{[11]}_2&\Rightarrow& {\bf H}^{[21]}{\bf V}^{[11]}_1\gamma^{[1]}_1+{\bf H}^{[21]}{\bf V}^{[11]}_2\\
u^{[12]}_2&\Rightarrow&{\bf H}^{[22]}{\bf V}^{[12]}_1\gamma^{[2]}_1+{\bf H}^{[22]}{\bf V}^{[12]}_2
\end{eqnarray}
Thus we would like
\begin{eqnarray}\label{eqn:null2}
\left[{\bf H}^{[21]}{\bf V}^{[11]}_1 ~~~ {\bf H}^{[21]}{\bf V}^{[11]}_2 ~~~ {\bf H}^{[22]}{\bf V}^{[12]}_1~~~{\bf H}^{[22]}{\bf V}^{[12]}_2\right]_{3\times 4}
\left[\begin{array}{c}
\gamma^{[1]}_1\\
1\\
-\delta\gamma^{[2]}_1\\
-\delta
\end{array}
\right]={\bf 0}_{3\times 1}
\end{eqnarray}
As before, we choose the values of $\gamma_1^{[1]}, \gamma^{[2]}_1, \delta$ to satisfy the equation above, and thereby achieve interference alignment at Receiver 2.

Lastly, we need to check that the desired symbols are not aligned either with each other or with the interference by this choice of the $\gamma^{[\star]}_\star$ constants.
Consider Receiver 1, where the desired symbols $u^{[11]}_2, u^{[12]}_2$ arrive along directions:
\begin{eqnarray}
u^{[11]}_2&\Rightarrow&{\bf H}^{[11]}{\bf V}^{[11]}_1\gamma^{[1]}_1+ {\bf H}^{[11]}{\bf V}^{[11]}_2\\
u^{[12]}_2&\Rightarrow&{\bf H}^{[12]}{\bf V}^{[12]}_1\gamma^{[2]}_1+{\bf H}^{[12]}{\bf V}^{[12]}_2
\end{eqnarray}
and the aligned interference arrives along the direction:
\begin{eqnarray}
{\bf H}^{[11]}{\bf V}^{[21]}_1\gamma^{[1]}_2+ {\bf H}^{[11]}{\bf V}^{[21]}_2 
\end{eqnarray}
Thus, for desired symbols to be resolvable from the interference, we need the following $3\times 3$ matrix to have full rank:
\begin{eqnarray}
M_1&=&\left[{\bf H}^{[11]}{\bf V}^{[11]}_1\gamma^{[1]}_1+ {\bf H}^{[11]}{\bf V}^{[11]}_2 ~~~ {\bf H}^{[12]}{\bf V}^{[12]}_1\gamma^{[2]}_1+{\bf H}^{[12]}{\bf V}^{[12]}_2~~~~{\bf H}^{[11]}{\bf V}^{[21]}_1\gamma^{[1]}_2+ {\bf H}^{[11]}{\bf V}^{[21]}_2 \right]
\end{eqnarray}
Similarly we can define the $3\times 3$ matrix $M_2$ that also needs to be full rank for Receiver 2 to be able to obtain his desired symbols. Both conditions can be stated together in the following form:
\begin{eqnarray}
\det(M_1)\det(M_2)&\neq &0
\end{eqnarray}
However, both $\det(M_1)$ and $\det(M_2)$ correspond to  polynomials in ${\bf V}^{[**]}, {\bf H}^{[**]}$ (the $\gamma$ can be evaluated in terms of ${\bf V}, {\bf H}$), and therefore, so does the product $\det(M_1)\det(M_2)$. Note that ${\bf V}, {\bf H}$ are picked independently over complex numbers. Therefore the polynomial $\det(M_1)\det(M_2)$ is either identically the zero polynomial, or it is non-zero almost surely for all realizations of ${\bf V}, {\bf H}$.   To prove that it is non-zero almost surely, it suffices to show that it is not the zero polynomial, which in turn is established if there exists any non-zero evaluation of $\det(M_1)\det(M_2)$. Indeed this is easily verified by a numerical example. Since such examples are easy to find (almost all choices work fine)  we will omit the explicit construction.

\section{Interference Channel with 3 Users}
Besides the $X$ channel \cite{MMK,Jafar_Shamai} and the MISO BC \cite{Weingarten_Shamai_Kramer}, the interference channel with more than 2 users is one of the earliest settings where interference alignment was first introduced \cite{Cadambe_Jafar_int}, and as such it is natural to ask if interference alignment is possible in this setting with only delayed CSIT? In this section we will study the interference channel with $3$ users. With full CSIT it is known that the $3$ user interference channel has $\frac{3}{2}$ DoF, which is higher than the $2$ user X channel's $\frac{4}{3}$ DoF. However, with delayed CSIT, because the transmitters are even more distributed in the $3$ user interference channel, it is not clear if it will continue to have a DoF advantage over the X channel. In this paper, we will show the achievability of only $\frac{9}{8}$ DoF for the $3$ user interference channel, which is less than the $\frac{8}{7}$ DoF that we are able to achieve for the $X$ channel with delayed CSIT. This is interesting because it shows that delayed CSIT is beneficial even in the $3$ user interference channel.   It is also interesting because it raises the question -- whether the 3 user IC in fact pays a greater price in DoF than the X channel for delayed CSIT.  The question remains wide open because the optimality of the schemes presented here is neither established nor conjectured.

\begin{figure}[!t]
\centering
\includegraphics[width=3.5in]{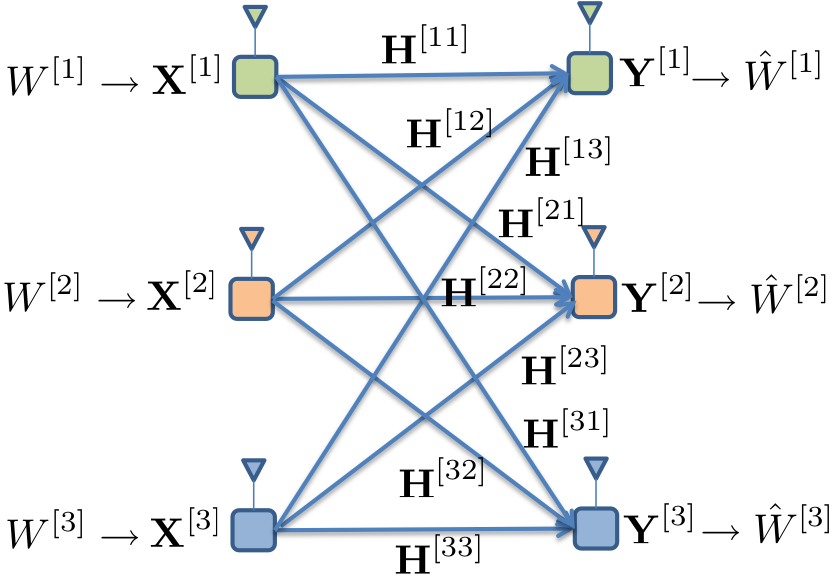}
\caption{Interference Channel with 3 Users}
\label{fig:3userIC}
\end{figure}

The 3 user interference channel, shown in Fig. \ref{fig:3userIC},  consists of transmitters 1,2,3, who wish to communicate independent messages $W^{[1]}, W^{[2]},W^{[3]}$ to receivers 1,2,3 respectively.  The assumptions regarding the delayed CSIT model are identical to the preceding sections, and the notation specific to the interference channel will become clear in the technical description of the proof.
The following theorem presents an achievability result for the DoF of the 3 user interference channel with delayed CSIT.
\begin{theorem}
The $3$ user interference channel with delayed CSIT, can achieve $\frac{9}{8}$ DoF almost surely.
\end{theorem}
\proof In order to show the achievability of $\frac{9}{8}$ DoF, we will consider a $8$ symbol extension of the channel. Each user will send $3$ information symbols over these $8$ channel uses. At each receiver, in addition to the $3$ desired symbols, there are $6$ interfering symbols. Since the total number of dimensions is only $8$, one of the $6$ interfering symbols must align within the vector space spanned by the remaining $5$, to leave $3$ signal dimensions free of interference where the desired signals can be projected. Since we are again dealing with delayed CSIT and distributed transmitters, we will again use the retrospective interference alignment scheme. However, in this section, for the sake of providing a richer understanding of the scheme, we will arrive at it in an alternative fashion.

{\bf Phase I: }
As stated earlier, we wish that the 6 interfering symbols should span no more than $5$ dimensions. Since we have no instantaneous channel knowledge, let us start by sending random linear combinations of the symbols from each transmitter. Since interference is allowed to fill up $5$ dimensions, we can send $5$ random linear combinations of the symbols from each transmitter over the first $5$ channel uses without exceeding the quota of $5$ dimensions that are allowed to be spanned by interference. This is the end of Phase I. No special effort has been made to align anything so far, and we have exhausted the number of dimensions allowed for interference at each receiver.

At this point, consider the signal seen by Receiver 1 (ignoring noise as usual).
\begin{eqnarray*}
{\bf Y}^{[1]}&=&{\bf H}^{[11]}\left[{\bf V}^{[1]}_1 ~~{\bf V}^{[1]}_2~~{\bf V}^{[1]}_3\right]\left[\begin{array}{c}u^{[1]}_1\\u^{[1]}_2\\u^{[1]}_3\end{array}\right]+{\bf H}^{[12]}\left[{\bf V}^{[2]}_1 ~~{\bf V}^{[2]}_2~~{\bf V}^{[2]}_3\right]\left[\begin{array}{c}u^{[2]}_1\\u^{[2]}_2\\u^{[2]}_3\end{array}\right]+{\bf H}^{[13]}\left[{\bf V}^{[3]}_1 ~~{\bf V}^{[3]}_2~~{\bf V}^{[3]}_3\right]\left[\begin{array}{c}u^{[3]}_1\\u^{[3]}_2\\u^{[3]}_3\end{array}\right]
\end{eqnarray*}
Here ${\bf V}^{[k]}_i$ are the $5\times 1$ precoding vectors, ${\bf Y}^{[k]}$ is the $5\times 1$ vector of received signals so far, ${\bf H}^{[kj]}$ is the $5\times 5$ diagonal channel matrix representing the i.i.d. variations of the channel coefficient from Transmitter $j$ to Receiver $k$. 

Consider the interference carrying vectors ${\bf V}^{[2]}_i, {\bf V}^{[3]}_i, i=1,2,3$, over the first $5$ channel uses. Since these six vectors are only in a five dimensional space, we can identify the $6\times 1$ null vector $\vec\alpha^{[1]}=[\alpha^{[1]}_1,\alpha^{[1]}_2,\cdots,\alpha^{[1]}_6]$ such that:
\begin{eqnarray}
\left[{\bf H}^{[12]}{\bf V}^{[2]}_1~~~{\bf H}^{[12]}{\bf V}^{[2]}_2~~~{\bf H}^{[12]}{\bf V}^{[2]}_3~~~{\bf H}^{[13]}{\bf V}^{[3]}_1~~~{\bf H}^{[13]}{\bf V}^{[3]}_2~~~{\bf H}^{[13]}{\bf V}^{[3]}_3\right]\vec\alpha^{[1]}={\bf 0}_{5\times 1}
\end{eqnarray}
Similarly, considering Receivers 2, and 3 respectively, we define null vectors $\vec{\alpha}^{[2]}, \vec\alpha^{[3]}$ so that
\begin{eqnarray}
\left[{\bf H}^{[21]}{\bf V}^{[1]}_1~~~{\bf H}^{[21]}{\bf V}^{[1]}_2~~~{\bf H}^{[21]}{\bf V}^{[1]}_3~~~{\bf H}^{[23]}{\bf V}^{[3]}_1~~~{\bf H}^{[23]}{\bf V}^{[3]}_2~~~{\bf H}^{[23]}{\bf V}^{[3]}_3\right]\vec\alpha^{[2]}={\bf 0}_{5\times 1}\\
\left[{\bf H}^{[31]}{\bf V}^{[1]}_1~~~{\bf H}^{[31]}{\bf V}^{[1]}_2~~~{\bf H}^{[31]}{\bf V}^{[1]}_3~~~{\bf H}^{[32]}{\bf V}^{[2]}_1~~~{\bf H}^{[32]}{\bf V}^{[2]}_2~~~{\bf H}^{[32]}{\bf V}^{[2]}_3\right]\vec\alpha^{[3]}={\bf 0}_{5\times 1}
\end{eqnarray}

{\bf Phase 2:} Phase 2 consists of the remaining $3$ channel uses. It is here that retrospective alignment will be used, based on the knowledge of the channel states from Phase I. No knowledge of channel states, not even delayed CSIT, will be used of the Phase-2 channels. Consider the $n^{th}$, $n=6,7,8$, transmission, i.e., any of the three transmissions of Phase 2. Suppose the transmitters choose to send the linear combinations:
\begin{eqnarray}
\mbox{Transmitter 1:}&\rightarrow& {\bf V}^{[1]}_1(n)u^{[1]}_1+{\bf V}^{[1]}_2(n)u^{[1]}_2+{\bf V}^{[1]}_3(n)u^{[1]}_3\\
\mbox{Transmitter 2:}&\rightarrow& {\bf V}^{[2]}_1(n)u^{[2]}_1+{\bf V}^{[2]}_2(n)u^{[2]}_2+{\bf V}^{[2]}_3(n)u^{[2]}_3\\
\mbox{Transmitter 3:}&\rightarrow& {\bf V}^{[3]}_1(n)u^{[3]}_1+{\bf V}^{[3]}_2(n)u^{[3]}_2+{\bf V}^{[3]}_3(n)u^{[3]}_3
\end{eqnarray}
The linear precoding coefficients ${\bf V}^{[k]}_i(n)$ can be chosen by the transmitters based on the delayed CSIT, i.e., the knowledge of the channel coefficients from Phase-I.

An important observation here is the following. In order to keep the interference contained in a 5 dimensional space at each receiver, the Phase-2 precoding coefficients must follow the same linear relationship as established in Phase-I. Mathematically, at Receiver 1:
\begin{eqnarray*}
\left[{\bf H}^{[12]}(n){\bf V}^{[2]}_1(n)~~~{\bf H}^{[12]}(n){\bf V}^{[2]}_2(n)~~~{\bf H}^{[12]}(n){\bf V}^{[2]}_3(n)~~~{\bf H}^{[13]}(n){\bf V}^{[3]}_1(n)~~~{\bf H}^{[13]}(n){\bf V}^{[3]}_2(n)~~~{\bf H}^{[13]}(n){\bf V}^{[3]}_3(n)\right]\vec\alpha^{[1]}=0
\end{eqnarray*}
and similarly at Receivers 2, 3:
\begin{eqnarray*}
\left[{\bf H}^{[21]}(n){\bf V}^{[1]}_1(n)~~~{\bf H}^{[21]}(n){\bf V}^{[1]}_2(n)~~~{\bf H}^{[21]}(n){\bf V}^{[1]}_3(n)~~~{\bf H}^{[23]}(n){\bf V}^{[3]}_1(n)~~~{\bf H}^{[23]}(n){\bf V}^{[3]}_2(n)~~~{\bf H}^{[23]}(n){\bf V}^{[3]}_3(n)\right]\vec\alpha^{[2]}=0\\
\left[{\bf H}^{[31]}(n){\bf V}^{[1]}_1(n)~~~{\bf H}^{[31]}(n){\bf V}^{[1]}_2(n)~~~{\bf H}^{[31]}(n){\bf V}^{[1]}_3(n)~~~{\bf H}^{[32]}(n){\bf V}^{[2]}_1(n)~~~{\bf H}^{[32]}(n){\bf V}^{[2]}_2(n)~~~{\bf H}^{[32]}(n){\bf V}^{[2]}_3(n)\right]\vec\alpha^{[3]}=0
\end{eqnarray*}

 Since the current channel states ${\bf H}^{[\star\star]}(n)$ are not known to the transmitters, the only way to guarantee the above equations for all current channel realizations is to choose ${\bf V}^{[k]}_i$ so that
\begin{eqnarray}
{\bf V}^{[1]}_1(n)\alpha^{[2]}_1+{\bf V}^{[1]}_2(n)\alpha^{[2]}_2+{\bf V}^{[1]}_3(n)\alpha^{[2]}_3=0\label{eq:v11}\label{eq:alignfirst}\\
{\bf V}^{[1]}_1(n)\alpha^{[3]}_1+{\bf V}^{[1]}_2(n)\alpha^{[3]}_2+{\bf V}^{[1]}_3(n)\alpha^{[3]}_3=0\label{eq:v12}\\
{\bf V}^{[2]}_1(n)\alpha^{[1]}_1+{\bf V}^{[2]}_2(n)\alpha^{[1]}_2+{\bf V}^{[2]}_3(n)\alpha^{[1]}_3=0\\
{\bf V}^{[2]}_1(n)\alpha^{[3]}_4+{\bf V}^{[2]}_2(n)\alpha^{[3]}_5+{\bf V}^{[2]}_3(n)\alpha^{[3]}_6=0\\
{\bf V}^{[3]}_1(n)\alpha^{[1]}_4+{\bf V}^{[3]}_2(n)\alpha^{[1]}_5+{\bf V}^{[3]}_3(n)\alpha^{[1]}_6=0\\
{\bf V}^{[3]}_1(n)\alpha^{[2]}_4+{\bf V}^{[3]}_2(n)\alpha^{[2]}_5+{\bf V}^{[3]}_3(n)\alpha^{[2]}_6=0\label{eq:alignlast}
\end{eqnarray}

Now consider the precoding coefficients ${\bf V}^{[1]}_i(n), i=1,2,3$ that are to be chosen by Transmitter 1. Based on equations (\ref{eq:v11}), (\ref{eq:v12}), we can express ${\bf V}^{[1]}_2(n)$ and ${\bf V}^{[1]}_3(n)$ as linear functions of ${\bf V}^{[1]}_1(n)$, say 
\begin{eqnarray}
{\bf V}^{[1]}_2(n)&=&\frac{\left|\begin{array}{rr}-\alpha^{[2]}_1&\alpha^{[2]}_3\\-\alpha^{[3]}_1&\alpha^{[3]}_3\end{array}\right|}{\left|\begin{array}{rr}\alpha^{[2]}_2&\alpha^{[2]}_3\\\alpha^{[3]}_2&\alpha^{[3]}_3\end{array}\right|}{\bf V}^{[1]}_1(n)\\
%c_1{\bf V}^{[1]}_1(n)\\
{\bf V}^{[1]}_3(n)&=&\frac{\left|\begin{array}{rr}\alpha^{[2]}_2&-\alpha^{[2]}_1\\\alpha^{[3]}_2&-\alpha^{[3]}_1\end{array}\right|}{\left|\begin{array}{rr}\alpha^{[2]}_2&\alpha^{[2]}_3\\\alpha^{[3]}_2&\alpha^{[3]}_3\end{array}\right|}{\bf V}^{[1]}_1(n)
\end{eqnarray}

Thus, Transmitter 1 is forced to send:
\begin{eqnarray}
{\bf X}^{[1]}(n)&=& {\bf V}^{[1]}_1(n)u^{[1]}_1+{\bf V}^{[1]}_2(n)u^{[1]}_2+{\bf V}^{[1]}_3(n)u^{[1]}_3\\
&=&c\underbrace{\left((\alpha^{[2]}_2\alpha^{[3]}_3-\alpha^{[2]}_3\alpha^{[3]}_2)u^{[1]}_1+(\alpha^{[2]}_3\alpha^{[3]}_1-\alpha^{[3]}_3\alpha^{[2]}_1)u^{[1]}_2+(\alpha^{[2]}_1\alpha^{[3]}_2-\alpha^{[2]}_2\alpha^{[3]}_1)u^{[1]}_3\right)}_{s^{[1]}}\nonumber
\end{eqnarray}
where $c$ is any constant, and without loss of generality we can set it to unity. The new information variable
\begin{eqnarray}
s^{[1]}=(\alpha^{[2]}_2\alpha^{[3]}_3-\alpha^{[2]}_3\alpha^{[3]}_2)u^{[1]}_1+(\alpha^{[2]}_3\alpha^{[3]}_1-\alpha^{[3]}_3\alpha^{[2]}_1)u^{[1]}_2+(\alpha^{[2]}_1\alpha^{[3]}_2-\alpha^{[2]}_2\alpha^{[3]}_1)u^{[1]}_3
\end{eqnarray}
 is precisely our Phase-2 variable, available only to Transmitter 1, and composed of variables only available to Transmitter 1. Thus, in Phase-2, even though the transmitter has three information symbols to send, it can only send scaled versions of the same effective scalar symbol $s^{[1]}$ in order to keep the interference aligned within $5$ dimensions at each receiver. Similarly, we can define the effective variables $s^{[2]}$ and $s^{[3]}$ to be sent by transmitters $2$ and $3$ over Phase 2. 
 
\begin{eqnarray}
s^{[2]}=(\alpha^{[1]}_2\alpha^{[3]}_6-\alpha^{[1]}_3\alpha^{[3]}_5)u^{[2]}_1+(\alpha^{[3]}_4\alpha^{[1]}_3-\alpha^{[3]}_6\alpha^{[1]}_1)u^{[2]}_2+(\alpha^{[1]}_1\alpha^{[3]}_5-\alpha^{[1]}_2\alpha^{[3]}_4)u^{[2]}_3\\
s^{[3]}=(\alpha^{[1]}_5\alpha^{[2]}_6-\alpha^{[1]}_6\alpha^{[2]}_5)u^{[3]}_1+(\alpha^{[1]}_6\alpha^{[2]}_4-\alpha^{[2]}_6\alpha^{[1]}_4)u^{[3]}_2+(\alpha^{[1]}_4\alpha^{[2]}_5-\alpha^{[2]}_4\alpha^{[1]}_5)u^{[3]}_3
\end{eqnarray}
 
Since there are only $3$ Phase-2 symbols, and there are $3$ channel uses, the operation over Phase-2 can be simply interpreted as each transmitter repeating its own effective information symbol, so that the channel variations provide each receiver with a different linear combination of the $3$ effective Phase-2 symbols each time, so that at the end of Phase-2, each receiver is able to decode all three symbols $s^{[1]}, s^{[2]}, s^{[3]}$. 

Thus, we have completely determined the resulting precoding vectors sent over the $8$ symbols. Putting everything together, the transmitted symbols from e.g., Transmitter 1 are shown below:

\begin{eqnarray}
\left[\begin{array}{c}
{\bf X}^{[1]}(1)\\
{\bf X}^{[1]}(2)\\
{\bf X}^{[1]}(3)\\
{\bf X}^{[1]}(4)\\
{\bf X}^{[1]}(5)\\
{\bf X}^{[1]}(6)\\
{\bf X}^{[1]}(7)\\
{\bf X}^{[1]}(8)
\end{array}\right]&=&
\left[\begin{array}{ccc}
{\bf V}^{[1]}_1(1)&{\bf V}^{[1]}_2(1)&{\bf V}^{[1]}_3(1)\\
{\bf V}^{[1]}_1(2)&{\bf V}^{[1]}_2(2)&{\bf V}^{[1]}_3(2)\\
{\bf V}^{[1]}_1(3)&{\bf V}^{[1]}_2(3)&{\bf V}^{[1]}_3(3)\\
{\bf V}^{[1]}_1(4)&{\bf V}^{[1]}_2(4)&{\bf V}^{[1]}_3(4)\\
{\bf V}^{[1]}_1(5)&{\bf V}^{[1]}_2(5)&{\bf V}^{[1]}_3(5)\\
\alpha^{[2]}_2\alpha^{[3]}_3-\alpha^{[2]}_3\alpha^{[3]}_2&
\alpha^{[2]}_3\alpha^{[3]}_1-\alpha^{[3]}_3\alpha^{[2]}_1&
\alpha^{[2]}_1\alpha^{[3]}_2-\alpha^{[2]}_2\alpha^{[3]}_1\\
\alpha^{[2]}_2\alpha^{[3]}_3-\alpha^{[2]}_3\alpha^{[3]}_2&
\alpha^{[2]}_3\alpha^{[3]}_1-\alpha^{[3]}_3\alpha^{[2]}_1&
\alpha^{[2]}_1\alpha^{[3]}_2-\alpha^{[2]}_2\alpha^{[3]}_1\\
\alpha^{[2]}_2\alpha^{[3]}_3-\alpha^{[2]}_3\alpha^{[3]}_2&
\alpha^{[2]}_3\alpha^{[3]}_1-\alpha^{[3]}_3\alpha^{[2]}_1&
\alpha^{[2]}_1\alpha^{[3]}_2-\alpha^{[2]}_2\alpha^{[3]}_1
\end{array}\right]\left[\begin{array}{c}u^{[1]}_1\\u^{[1]}_2\\u^{[1]}_3\end{array}\right]
\end{eqnarray}
The first 5 channel uses correspond to Phase 1. All these precoding coefficients ${\bf V}^{[1]}_i(n)$ are chosen independently, randomly, before the beginning of communication and with no knowledge of CSIT. The last three channel uses correspond to Phase 2, and can be easily seen to be repetitions of the Phase-2 variable $s^{[1]}$. The transmitted symbols for all other transmitters can be described similarly.  Interference alignment is accomplished because this choice of precoding vectors satisfies equations (\ref{eq:alignfirst})-(\ref{eq:alignlast}).

Keeping the interference aligned within $5$ dimensions at each receiver, allows the receiver to null out the $5$ interference dimensions and recover the $3$ desired symbols from the remaining $3$ dimensions from the overall $8$ dimensional vector space. Once again, while the construction above guarantees that interference is restricted within $5$ dimensions, one must also show that the desired signal vectors are not aligned within the interference or aligned among themselves. This is proven as before, by constructing the $8\times 8$ matrix consisting of 3 desired signal vectors and 5 interference vectors that span the interference space received at each receiver, and showing that the determinant of this matrix, which is equivalent to a polynomial in Phase 1 variables, is not identically a zero polynomial. While we have established this through numerical evaluations, the details of an explicit numerical example are omitted here (because almost all examples work).

\section{Delayed Output Feedback}
So far we assumed that the only feedback available to the transmitters consists of delayed CSIT. Another commonly studied model for feedback is channel \emph{output} feedback (without explicitly providing the CSI). In this section we study the X channel and the $3$ user interference channel with delayed \emph{output} feedback, i.e., the channel output is available to the transmitters only after the channel state associated with the observed output is no longer current. While delayed CSIT created difficulties because of the transmitters' inability to reconstruct the previously received linear combinations of undesired received symbols because of the distributed nature of the information, delayed output feedback automatically provides the transmitters with information that has been previously observed at one of the receivers. Retransmitting this information provides the transmitters an opportunity to provide new observations to the receivers who desire this information, while allowing the receivers who have already observed this interference to cancel it entirely. In this sense, delayed output feedback allows a direct extension of the alignment techniques explored in \cite{Maddah_Tse}. 
\subsection{X Channel with Delayed Output Feedback}
\begin{theorem}
The  X channel with delayed output feedback can achieve $\frac{4}{3}$ DoF almost surely.
\end{theorem}
\proof
In light of the earlier discussion on the vector BC, it is easily seen that the X channel, with only delayed channel output feedback can easily achieve the $\frac{4}{3}$ DoF outer bound. Since the achievable scheme is essentially the same as the schemes studied  \cite{Jafar_corr} and \cite{Maddah_Tse}, we only present a brief description.
\begin{itemize}
\item Send two symbols to each receiver over three time slots to achieve $\frac{4}{3}$ DoF.
\item In the first time slot, each transmitter sends its own symbol intended for Receiver 1. Receiver 1 observes a linear combination (along with noise) of the two desired symbols, while Receiver 2 sees a linear combination of undesired symbols (and noise). 
\item The second time slot is similar to the first time slot, except the information symbols transmitted are for User 2.
\item In the third time slot, the transmitters send a superposition of the previous undesired outputs. This is possible due to delayed output feedback as long as each undesired output signal is available to one of the transmitters.
\end{itemize}
The third time slot provides each receiver with a second linear combination of desired symbols while the interfering undesired information is cancelled because it has been received previously. Note that we are assuming that each receiver has \emph{global} channel knowledge, i.e., it knows not only the channels associated with itself but also the other receivers' channels as well. Further we are once again ignoring noise in this discussion because, as stated earlier, for such linear beamforming schemes, noise does not affect the DoF.
\subsection{3 User Interference Channel}
The following theorem presents an achievability result for the 3 user interference channel.
\begin{theorem}
The  3 user interference channel with delayed output feedback available from each receiver to only its corresponding transmitter can achieve $\frac{6}{5}$ DoF almost surely.
\end{theorem}
{\it Remark:} While it is remarkable that the achievability results presented in this work for both the X channel and the 3 user interference channel under delayed output feedback correspond to higher DoF than with delayed CSIT, it should be noted that these are only achievability results and in the absence of outer bounds it is not possible to make categorical comparisons between the two settings.

\proof

In order to achieve $\frac{6}{5}$ DoF we will operate over a $5$ channel-use block. Each user will communicate two coded information symbols over these $5$ channel uses using linear schemes that can be simply seen as swapping output symbols to help resolve desired signals \cite{Maddah_Tse}. A summary of the transmission scheme is described below.
\begin{enumerate}
\item Over the first time slot, Transmitter 1 sends its first information symbol $u^{[1]}_1$ and Transmitter 2 simultaneously sends its first information symbol $u^{[2]}_1$. Ignoring noise, the received signals are described below:
\begin{eqnarray}
\mbox{Receiver 1: }Y^{[1]}(1)&=&H^{[11]}(1)u^{[1]}_1+H^{[12]}(1)u^{[2]}_1\\
\mbox{Receiver 2: }Y^{[2]}(1)&=&H^{[21]}(1)u^{[1]}_1+H^{[22]}(1)u^{[2]}_1\\
\mbox{Receiver 3: }Y^{[3]}(1)&=&H^{[31]}(1)u^{[1]}_1+H^{[32]}(1)u^{[2]}_1
\end{eqnarray}
\item  Over the second time slot, Transmitter 1 sends its second information symbol $u^{[1]}_2$ and Transmitter 3 simultaneously sends its first information symbol $u^{[3]}_1$. Ignoring noise, the received signals are described below:
\begin{eqnarray}
\mbox{Receiver 1: }Y^{[1]}(2)&=&H^{[11]}(2)u^{[1]}_1+H^{[13]}(2)u^{[3]}_1\\
\mbox{Receiver 2: }Y^{[2]}(2)&=&H^{[21]}(2)u^{[1]}_1+H^{[23]}(2)u^{[3]}_1\\
\mbox{Receiver 3: }Y^{[3]}(2)&=&H^{[31]}(2)u^{[1]}_1+H^{[33]}(2)u^{[3]}_1
\end{eqnarray}
\item Over the third time slot, Transmitter 2 sends its second information symbol $u^{[2]}_2$ and Transmitter 3 simultaneously sends its second information symbol $u^{[3]}_2$. Ignoring noise, the received signals are described below:
\begin{eqnarray}
\mbox{Receiver 1: }Y^{[1]}(3)&=&H^{[11]}(3)u^{[2]}_2+H^{[12]}(3)u^{[3]}_2\\
\mbox{Receiver 2: }Y^{[2]}(3)&=&H^{[21]}(3)u^{[2]}_2+H^{[22]}(3)u^{[3]}_2\\
\mbox{Receiver 3: }Y^{[3]}(3)&=&H^{[31]}(3)u^{[2]}_2+H^{[32]}(3)u^{[3]}_2
\end{eqnarray}
\item Over the fourth time slot, Transmitter 3 retransmits $Y^{[3]}(1)$ and Transmitter 2 retransmits $Y^{[2]}(2)$.
\item Over the fifth time slot, $Y^{[3]}(1)$ and $Y^{[1]}(3)$ are retransmitted from Transmitters $3$ and $1$ respectively.
\end{enumerate}
Next we explain how every receiver has enough information to recover its two desired symbols. 

{\bf Receiver 1:} Consider Receiver 1. From the linear combination of $Y^{[3]}(1)$ and $Y^{[1]}(3)$ received over the fifth symbol, this receiver is able to remove $Y^{[1]}(3)$ which it has previously received, to obtain $Y^{[3]}(1)$. Combining $Y^{[1]}(1)$ and $Y^{[3]}(1)$ the receiver has enough information to resolve the first received symbol $u^{[1]}_1$.

Further, from the linear combination of $Y^{[3]}(1), Y^{[2]}(2)$ received over the fourth symbol, the receiver removes $Y^{[3]}(1)$ to obtain a clean $Y^{[2]}(2)$. Combining $Y^{[2]}(2)$ with $Y^{[1]}(2)$, the receiver is able to resolve the second desired symbol $u^{[1]}_2$.

{\bf Receiver 2:} Consider Receiver 2. From the linear combination of  $Y^{[3]}(1), Y^{[2]}(2)$ received over the fourth symbol, this receiver is able to remove $Y^{[2]}(2)$ which it has previously received, to obtain $Y^{[3]}(1)$. Combining $Y^{[2]}(1)$ and $Y^{[3]}(1)$ the receiver has enough information to resolve the first desired symbol $u^{[2]}_1$.

Further, from the linear combination of $Y^{[3]}(1)$ and $Y^{[1]}(3)$ received over the fifth channel use, the receiver removes $Y^{[3]}(1)$ to obtain a clean $Y^{[1]}(3)$. Combining $Y^{[1]}(3)$ with $Y^{[2]}(3)$, the receiver is able to resolve the second desired symbol $u^{[2]}_2$.

{\bf Receiver 3:} Consider Receiver 3. From the linear combination of  $Y^{[3]}(1)$ and  $Y^{[2]}(2)$ received over the fourth symbol, this receiver is able to remove $Y^{[3]}(1)$ which it has previously received, to obtain $Y^{[2]}(2)$. Combining $Y^{[2]}(2)$ and $Y^{[3]}(2)$ the receiver has enough information to resolve the first received symbol $u^{[3]}_1$.

Further, from the linear combination of $Y^{[1]}(3), Y^{[3]}(1)$ received over the fifth symbol, the receiver removes $Y^{[3]}(1)$ to obtain a clean $Y^{[1]}(3)$. Combining $Y^{[1]}(3)$ with $Y^{[3]}(3)$, the receiver is able to resolve the second desired symbol $u^{[3]}_2$.

Thus, all symbols are resolved and $\frac{6}{5}$ DoF are achieved on the 3 user interference channel.
\section{Conclusion}
We explored similarities, differences, and the apparent difficulties in achieving interference alignment with channel uncertainty at the transmitters based on recent works that assume two different channel uncertainty models -- staggered block fading and delayed CSIT. While there are many shared aspects that allow the schemes to be translated from one setting to another for many cases, overall the two settings are indeed fundamentally different and face different challenges. In particular, the delayed CSIT setting appears to be more sensitive to whether the transmitters are co-located or distributed, unlike previous results where both for compound channels and suitably staggered block fading models the two were found to be equivalent from a DoF perspective. While the $2$ user MISO BC with delayed CSIT easily achieves the outer bound of $\frac{4}{3}$, it is not known if the same DoF can be achieved on the X channel, i.e., without cooperation between transmitters. We were able to show that delayed CSIT is still useful in the X channel from a DoF perspective, as one can achieve $\frac{8}{7}$ DoF. The achievability was shown here using an interesting new scheme that we call retrospective interference alignment. While the scheme operates in two phases, and with two layers of variables as the scheme proposed in \cite{Maddah_Tse}, the novelty of retrospective alignment appears in the construction of auxiliary (layer - 2 in the terminology of \cite{Maddah_Tse}) variables that aid in the alignment of the previously transmitted information symbols based on only the information symbols available to each transmitter. The same scheme was used to prove the achievability of 9/8 DoF for the 3 user interference channel with delayed CSIT. We also found that the X channel and the 3 user interference channel can achieve $4/3$ and $6/5$ DoF respectively when delayed output feedback is available to the transmitters. It is remarkable that with perfect and instantaneous CSIT, output feedback does not increase DoF for X networks or interference channels \cite{Cadambe_Jafar_XFB}. 

Another issue that is of both theoretical and practical interest is the feedback rate. The delayed feedback models discussed in this work could be seen as essentially noiseless, infinite capacity feedback links. It is also clear that the benefits of feedback will be retained if the accuracy of feedback information is scaled appropriately with SNR \cite{Jindal, Caire_Jindal_Shamai}. An interesting question to make further progress in this direction would be to explore how a delayed feedback link whose capacity is itself limited in DoF becomes a bottleneck on the forward channel capacity. In conclusion, the results reported here only scratch the surface and much more remains to be done in order to understand the true potential for interference alignment in delayed feedback settings.

\bibliography{Thesis}

\end{document}